\documentclass[aps,pre,twocolumn,superscriptaddress,reprint]{revtex4-1}
\usepackage{amsmath}
\usepackage{amsfonts}
\usepackage{amssymb}
\usepackage{bm}
\usepackage[dvipdfmx]{graphicx}
\usepackage[dvipdfmx]{color}

\newcommand{\ket}[1]{|#1\rangle}
\newcommand{\bra}[1]{\langle#1|}

\begin{document}

\title{Steady state distributions of network of degenerate optical parametric oscillators in solving combinatorial optimization problems}

\author{Ryoji Miyazaki}
\author{Masayuki Ohzeki}
\affiliation{Graduate School of Information Science, Tohoku University, Sendai 980-8579, Japan}

\date{\today}

\begin{abstract}
We investigate network of degenerate optical parametric oscillators (DOPOs) 
as a model of the coherent Ising machine, 
an architecture for solving Ising problems.
The network represents the interaction in the Ising model, 
which is a generalization of a previously proposed one for the two-DOPO case.
Dynamics of the DOPOs is described by the Fokker-Planck equation in the positive $P$ representation.
We obtain approximate steady state distributions for arbitrary Ising problems under some ansatz.
Using the method of statistical mechanics, 
we analytically demonstrate that 
the most probable states in a particular range of the parameters correspond to the true optimal states for two rather simple problems, 
i.e., fully-connected ferromagnetic coupling without/with binary random fields.
In particular, for the random-field problem, 
the distribution correctly detects the phase transition 
that occurs in the target Ising model with varying the magnitude of the fields.
\end{abstract}

\pacs{...}

\maketitle

\section{Introduction}

Combinatorial optimization problems that have many applications can be translated into problems to find ground states of the Ising model~\cite{A.Lucas2014}.
This formulation motivates us to develope machines speciallizing in the search for the ground states.
A well-known example of such machines is the hardware devices provided by D-Wave Systems Inc.~\cite{M.Johnson2011}.
The devices implement quantum annealing~\cite{T.Kadowaki1998, J.Brooke1999, G.Santoro2002, A.Das2008} 
(or adiabatic quantum computation~\cite{E.Farhi2000, T.Albash2018}), 
a heuristic which harnesses quantum effects, 
instead of themal effects in simulated annealing~\cite{S.Kirkpatrick1983}, 
to search for the ground states.
It has had an impact to impelement quantum annealing directly using physics of artificial spins.
Other machines for such a purpose, 
solving problems in terms of the Ising model, 
have also been proposed and actually developed, 
which utilize, or are inspired by, interaction and dynamics in underlying physical phenomena~\cite{J.Britton2012, A.Marandi2014, T.Inagaki2016Oct, P.McMahon2016, I.Mahboob2016, M.Yamaoka2016, H.Goto2016, S.Nigg2017, S.Puri2017, B.Sutton2017, S.Tsukamoto2017, O.Kyriienko2018arXiv}.

The coherent Ising machine (CIM) is such a machine
based on network of degenerate optical parametric oscillators (DOPOs)~\cite{Z.Wang2013, A.Marandi2014, T.Inagaki2016Oct, P.McMahon2016}.
The degree of freedom we utilize as an Ising spin is phase of the signal field of a DOPO.
The signal field is amplified by the pump field via interaction in a nonlinear optical crystal in a cavity~\cite{A.Yariv2006, H.Carmichael2002, *H.Carmichael2008}.
Above the threshold of the pump field, 
the phase of the signal field bifurcates.
The phase difference from the pump field takes either 0 or $\pi$ at random, 
which encodes an Ising spin for the CIM.
Interaction of DOPOs makes correlation in their phases~\cite{A.Marandi2012Mar} as interaction of Ising spins.
Network composed of configurable interactions of DOPOs with the pump field around the threshold is thus expected to represent the lowest energy states of the corresponding Ising model~\cite{Z.Wang2013}.

Such network in the CIM was experimentally constructed with a system of time-multiplexed DOPO pulses in a ring cavity~\cite{A.Marandi2014, T.Inagaki2016Jun, K.Takata2016}.
Optical coupling of the pulses is realized by delay lines connected to the main ring.
The length of each delay line is tuned to be an integer multiple of the pulse-repetition period so that a fraction of a pulse taking a detour via a delay line interacts to another one running after it.
This system almost surely found the ground states of some Ising models~\cite{A.Marandi2014, T.Inagaki2016Jun, K.Takata2016}, 
but connection of spins is limited because of difficulty in making arbitrary graph structure under this scheme.
For instance, 
the regular graphs of degree-$k$ need $k$ delay lines.
To deal with this issue, 
the CIM with aid of field-programmable gate array (FPGA) has been developed, 
where FPGA manages the coupling of the pulses running in a ring cavity~\cite{T.Inagaki2016Oct, P.McMahon2016}.
This type of CIM can treat up to 2000 Ising spins with all-to-all couplings, 
exhibiting faster convergence to states for comparable or lower Ising energy than finely tuned simulated annealing running on a CPU.
Exploiting FPGA, 
the CIM does not show extreme decrease in the performance even for problems with fully-connected graphs.
This feature contributes to the CIM's advantage 
when we compare the CIM with the machines of D-Wave systems Inc.~\cite{R.Hamerly2018arXiv, C.McGeoch2018arXiv} 
adopting the so-called ``Chimera graph''~\cite{P.Bunyk2014, V.Choi2008, V.Choi2011, J.Cai2014arXiv}.
On the other hand, 
under the use of FPGA, 
the effect of dynamics of the DOPOs for the performance is murky~\cite{A.King2018arXiv}.
It is also obscure whether properties of DOPO, in particular, as quantum light are exploited to accelerate finding the solution.

We have needed theoretical description of the CIM to estimate its efficiency to solve Ising problems and also to clarify the dominant property for the efficiency.
Dynamics in the CIM based on quantum mechanics has been investigated~\cite{K.Takata2015, D.Maruo2016, T.Shoji2017, A.Yamamura2017}.
The detailed analysis, however, sufferred from huge comutational cost, 
and the system size was limited. 
The CIM for large-size problems has been numerically simulated by the semiclassical counterpart~\cite{T.Inagaki2016Oct, P.McMahon2016, Y.Haribara2017}.
The numerical simulation is a powerful tool to gain insight to the CIM. 
We, however, cannot conclude the performance of the CIM in general on the basis of behavior for particular instances observed with the numerical simulations.

For estimating the efficiency of the CIM without the dynamics simulations, 
it is a good strategy to find a factor 
that deteremines whether the CIM could find the correct solutions of a problem and 
that scales the computaion time required to solve it.
For quantum annealing, such a factor is the minimal energy gap between the ground state and the first excited one of problem Hamiltonian in an annealing passage~\cite{T.Albash2018}. 
The problem-size dependence of the minimal energy gap is often used to represent the difficulty of the problem for quantum annealing.
This formulation is generalized into open quantum systems, 
where the gap of eigenvalues of Liouvillian, instead of Hamiltonian, takes the role~\cite{L.Venuti2016}.
Considering the CIM as an annealing machine, 
we would obtain the factor along the argument similar to that for other annealing algorithms~\cite{S.Geman1984, H.Nishimori1998, T.Albash2018, L.Venuti2016}.
To this end, 
we first need the instantaneous steady state or its distributions of the CIM for the parameters.
We additionally have to check that there exist values of the parameters 
for which the steady state gives ground states of the target Ising model, 
since the cost function in the CIM, 
presented below, 
does not agree with the Hamiltonian of the Ising model.
Then we will move to a stage at which 
we discuss the factor and also the schedule to surely reach the optimal state for the problem by adiabatical evolution.

In this paper, 
we explore the steady state distributions of a model of the CIM.
Our aim is to clarify properties of a large number of DOPOs in solving combinatorial optimization problems, in particular, in the long time limit, 
where the steady state distributions are possibly realized.
The model does not include any component corresponding to the FPGA and the delay lines 
but is just simple network of DOPOs optically interacting with each other.
We deduce the Fokker-Planck equation 
describing the dynamics of DOPOs in the network.
Under some ansatz, 
we show the approximate steady state distribution.
Statistical-mechanical approach enables us to investigate the distributions for large-size problems.
We then examine the distributions and the most probable states for rather simple problems.

\section{Model}

We investigate network of DOPOs interacting with each other via mutual-injection paths as a theoretical model of the CIMs.
Our model is a generalization of the model for two DOPOs previously proposed~\cite{K.Takata2015}, 
which is also a generalization of the single DOPO model~\cite{P.Drummond1981} by adding the interaction of the DOPOs via a mutual injection path.
In the two-DOPO model the signal fields of a frequency $\omega$, 
which are used to represent Ising variables later, 
are assumed to be highly confined in each cavity and the path.
The spatial-phase factor $e^{ikd}$ of the bosonic operator in the injection path is considered, 
where $k$ is the wave number for the signal mode, 
and $d$ is the length of the path.
The mutual injection leads to in-phase couplings, i.e., the ferromagnetic coupling in terms of spin systems, if $e^{ikd} = 1$ and out of phase, i.e., the antiferromagnetic one, if $e^{ikd} = -1$.
The signal field is amplified in a nonlinear crystal via the interaction with the pump mode of a frequency $\omega_p = 2\omega$.
The pump mode is excited by the classical, driving field entering each DOPO.
The driving field is also used as the phase reference.
Above the threshold of the pump strength, 
the bifurcation of quadrature amplitude of each signal field is observed.
The sign of the quadrature amplitude encodes an Ising spin.
The DOPOs output the configuration of Ising spins according to their coupling in which the target Ising Hamiltonian is embedded.

We generalize the two-DOPO model into a system of $N$ DOPOs.
We assume that each pair of two DOPOs in the system interacts as in the above two-DOPO model.
While there exists similar generalization to the one-dimensional ring network consisting of $N$ DOPOs~\cite{D.Maruo2016}, 
we here treat almost arbitrary network as well as the implementation of the Zeeman term in the Ising Hamiltonian.
The Hamiltonian for our model is written as
\begin{equation}
H = \sum_{j=1}^N H^{(j)}_\text{DOPO} + \sum_{j<l} H^{(jl)}_\text{int} + \sum_{j=1}^N H^{(j)}_{Z} + H_\text{res}.
\end{equation}
The first sum includes~\cite{P.Drummond1981}
\begin{equation}
\begin{split}
H^{(j)}_\text{DOPO} 
=& \hbar\omega \hat{a}^{\dagger}_{sj} \hat{a}_{sj} + 2\hbar\omega \hat{a}^{\dagger}_{pj} \hat{a}_{pj}
 + i\hbar \frac{\kappa}{2} \left( \hat{a}^{\dagger 2}_{sj} \hat{a}_{pj} - \hat{a}^{\dagger}_{pj} \hat{a}^2_{sj}\right) \\
& + i\hbar \left( \epsilon_p \hat{a}^{\dagger}_{pj} e^{-2i\omega t} - \epsilon_p \hat{a}_{pj} e^{2i\omega t} \right), 
\end{split}
\end{equation}
where $\hat{a}_{sj}$ and $\hat{a}^{\dagger}_{sj}$ are the bosonic annihilation and creation operators, respectively, 
for the signal modes $j = 1, 2, \dots , N$, 
and $\hat{a}_{pj}$ and $\hat{a}^{\dagger}_{pj}$ are for the pump modes $j$.
The coupling constant of quadratic nonlinear interaction of the signal and pump modes is denoted by $\kappa$.
The pump mode is excited by the real driving field $\epsilon_p$ of a frequency $\omega_d = \omega_p = 2\omega$.
The Ising variable for discrete optimization problems is encoded in the sign of the in-phase amplitude of the signal mode $(\hat{a}_{sj}+\hat{a}_{sj}^\dagger)/2$~\cite{T.Inagaki2016Oct, P.McMahon2016, Z.Wang2013, A.Marandi2014, K.Takata2015}, 
which can be observed via homodyne detection.
We consider the beam-splitter interaction Hamiltonian between the signal modes in the cavity and the injection path 
to tune the interactions between different DOPOs, 
\begin{equation}
\begin{split}
H^{(jl)}_\text{int} 
= \hbar\omega \hat{a}^{\dagger}_{cjl} \hat{a}_{cjl}
+ & i\hbar \zeta_{jl}\Big( \hat{a}_{cjl}\hat{a}_{sj}^\dagger - \hat{a}_{cjl}^\dagger\hat{a}_{sj} \\
&+ \hat{a}_{sl}\hat{a}_{cjl}^\dagger e^{-i\theta_{jl}} - \hat{a}_{sl}^\dagger\hat{a}_{cjl}e^{i\theta_{jl}} \Big), 
\end{split}
\end{equation}
where the signal modes in the injection paths for $j$ and $l$ are denoted by $\hat{a}_{cjl}$ and $\hat{a}^{\dagger}_{cjl}$, 
and $\zeta_{jl}$ denotes the interaction coefficient of the signal modes and the injection-path mode for the path between cavities $j$ and $l$.
Phase $\theta_{jl}$ is equal to $kd_{jl}$, 
where $d_{jl}$ is the path length between the cavities.
Hamiltonian $H^{(j)}_Z$ is for excitation of the signal mode by the real field $\epsilon_s$ of a frequency $\omega$ to tune the effect for the Zeeman term in the Ising Hamiltonian,
\begin{equation}
H^{(j)}_Z = i\hbar \left(\epsilon_{sj}\hat{a}_{sj}^\dagger e^{-i\omega t} - \epsilon_{sj}\hat{a}_{sj}e^{i\omega t}\right).
\end{equation}
We do not explicitly show $H_\text{res}$, 
which is a standard one for interaction with surroundings (reserviors)~\cite{H.Carmichael2002, P.Drummond1981, K.Takata2015}.

The master equation for the density operator $\hat{\rho}$ for the system, 
where the degrees of freedom of the reseviors are traced out, 
is obtained under standard approximations introduced to treat the reserviors~\cite{H.Carmichael2002, *H.Carmichael2008, H.Breuer2007}; 
the Born-Markov approximation and neglecting implicit interactions of the reserviors through the internal couplings in the system. 
We set the reservior at zero temperature to eliminate thermal effects.
Accoridingly, noises in dynamics will be derived only from quantum fluctuations.
The resulting master equation is
\begin{equation}
\begin{split}
\frac{d\hat{\rho}}{dt} 
= & \frac{1}{i\hbar}\left[ \sum_{j=1}^N H^{(j)}_\text{DOPO} + \sum_{j<l} H^{(jl)}_\text{int} + \sum_{j=1}^N H^{(j)}_{Z}, \hat{\rho} \right] \\
&+ \sum_{j=1}^N 2\gamma_s \left( \hat{a}_{sj}\hat{\rho}\hat{a}_{sj}^\dagger -\frac{1}{2} \left\{ \hat{a}_{sj}^\dagger\hat{a}_{sj}, \hat{\rho} \right\} \right) \\
&+ \sum_{j=1}^N 2\gamma_p \left( \hat{a}_{pj}\hat{\rho}\hat{a}_{pj}^\dagger -\frac{1}{2} \left\{ \hat{a}_{pj}^\dagger\hat{a}_{pj}, \hat{\rho} \right\} \right) \\
&+ \sum_{j<l}^N 2\gamma_c \left( \hat{a}_{cjl}\hat{\rho}\hat{a}_{cjl}^\dagger -\frac{1}{2} \left\{ \hat{a}_{cjl}^\dagger\hat{a}_{cjl}, \hat{\rho} \right\} \right), 
\end{split} \label{eq:master}
\end{equation}
where $\gamma_s$, $\gamma_p$, and $\gamma_c$ are coefficients for the decay of the signal, pump, and injection-path modes through dissipation, respectively.

We here utilize the positive $P$ representation~\cite{P.Drummond1980Jan} to analyze the master equation.
The density operator in the positive $P$ representaiton is expanded in terms of the coherent product states and a distribution function $P(\bm{\alpha}, \bm{\beta})$ as
\begin{equation}
\hat{\rho} = \int d^{4N}\bm{\alpha}d^{4N}\bm{\beta} P(\bm{\alpha},\bm{\beta}) \frac{\ket{\bm{\alpha}}\bra{\bm{\beta}^*}}{\langle\bm{\beta}^*|\bm{\alpha}\rangle}. \label{eq:def_positive-P}
\end{equation}
The $c$-number vector $\bm{\alpha}$ composed of $\alpha_{sj}$, $\alpha_{pj}$, and $\alpha_{cjl}$ for $j, l=1, \dots , N$ and $j<l$ represents the coherent product state $\ket{\bm{\alpha}} = \prod_{j=1}^N \ket{\alpha_{sj}}\ket{\alpha_{pj}}\prod_{j<l}\ket{\alpha_{cjl}}$, 
and $\bm{\beta}$ describes another one $\bra{\bm{\beta}^*} = \prod_{j=1}^N \bra{\beta^*_{sj}}\bra{\beta^*_{pj}}\prod_{j<l}\bra{\beta^*_{cjl}}$.
The distribution function $P(\bm{\alpha}, \bm{\beta})$ itself in the positive $P$ representation is not uniquely determined, 
but the normal-ordered average is calculated with any distribution $P(\bm{\alpha}, \bm{\beta})$ that satisfies Eq.~(\ref{eq:def_positive-P}).
The nonuniqueness allows the distribution to be real and positive~\cite{P.Drummond1980Jan, H.Carmichael2002, *H.Carmichael2008}, 
even when the density operator is composed of the superposition of different coherent states.
In this expression the average of in-phase amplitude $(\hat{a}_{sj} + \hat{a}_{sj}^\dagger)/2$ 	is computed as the average of $(\alpha_j + \beta_j)/2$ over the distribution $P(\bm{\alpha},\bm{\beta})$. 

Substituting Eq.~(\ref{eq:def_positive-P}) into Eq.~(\ref{eq:master}), 
we obtain the Fokker-Planck equation for the distribution $P(\bm{\alpha},\bm{\beta})$ 
through typical calculations for this representation~\cite{H.Carmichael2002, *H.Carmichael2008, K.Takata2015}, 
\begin{widetext}
\begin{equation}
\begin{split}
\frac{dP(\bm{\alpha},\bm{\beta})}{dt}
= \Bigg\{ \sum_{j=1}^N & \Bigg[ \frac{\partial}{\partial \alpha_{sj}} \left(\gamma_s \alpha_{sj} - \kappa\beta_{sj}\alpha_{pj} -\sum_{l(>j)}\zeta_{jl}\alpha_{cjl} + \sum_{l(<j)}\zeta_{lj}\alpha_{clj}e^{i\theta_{lj}} -\epsilon_{sj} \right) \\
&+ \frac{\partial}{\partial \beta_{sj}} \left(\gamma_s \beta_{sj} - \kappa\alpha_{sj}\beta_{pj} -\sum_{l(>j)}\zeta_{jl}\beta_{cjl} + \sum_{l(<j)}\zeta_{lj}\beta_{clj}e^{-i\theta_{lj}} -\epsilon_{sj} \right) \\
& + \frac{1}{2}\frac{\partial^2}{\partial\alpha_{sj}^2} \kappa \alpha_{pj} + \frac{1}{2}\frac{\partial^2}{\partial\beta_{sj}^2} \kappa \beta_{pj} 
+ \frac{\partial}{\partial\alpha_{pj}} \left(\gamma_p\alpha_{pj}-\epsilon_p+\frac{\kappa}{2}\alpha_{sj}^2\right) 
+ \frac{\partial}{\partial\beta_{pj}} \left(\gamma_p\beta_{pj}-\epsilon_p+\frac{\kappa}{2}\beta_{sj}^2\right) \Bigg] \\
+ \sum_{j<l} & \left\{ \frac{\partial}{\partial\alpha_{cjl}} \left[ \gamma_c\alpha_{cjl} + \zeta_{jl}\left(\alpha_{sj}-\alpha_{sl}e^{-i\theta_{jl}}\right)\right]+ \frac{\partial}{\partial\beta_{cjl}} \left[\gamma_c\beta_{cjl} + \zeta_{jl}\left(\beta_{sj}-\beta_{sl}e^{i\theta_{jl}}\right)\right] \right\} \Bigg\} P(\bm{\alpha},\bm{\beta}) 
\end{split}
\end{equation}
\end{widetext}
where we have taken the rotating frame with $\omega$ for the signal modes and $2\omega$ for the pump modes.

The Ito rule leads to the corresponding stochastic differential equations:
\begin{equation}
\begin{split}
d\alpha_{sj} 
=& \Bigg(-\gamma_s \alpha_{sj} + \kappa\beta_{sj}\alpha_{pj} \\
&+\sum_{l(>j)}\zeta_{jl}\alpha_{cjl} - \sum_{l(<j)}\zeta_{lj}\alpha_{clj}e^{i\theta_{lj}} +\epsilon_{sj} \Bigg)dt \\
&+\sqrt{\kappa\alpha_{pj}} dW_{\alpha_{sj}}(t),
\end{split} \label{eq:dalphasj}
\end{equation}
\begin{equation}
\begin{split}
d\beta_{sj} 
=&\Bigg(-\gamma_s \beta_{sj} + \kappa\alpha_{sj}\beta_{pj} \\
&+\sum_{l(>j)}\zeta_{jl}\beta_{cjl} - \sum_{l(<j)}\zeta_{lj}\beta_{clj}e^{-i\theta_{lj}} +\epsilon_{sj} \Bigg)dt \\
&+\sqrt{\kappa\beta_{pj}} dW_{\beta_{sj}}(t), 
\end{split} \label{eq:dbetasj} 
\end{equation} 
where $dW_{x}(t)$ is the standard Wiener increment for variable $x$.
Similarly, we obtain those for the pump modes,
\begin{align}
d\alpha_{pj} 
=&\left(-\gamma_p\alpha_{pj}+\epsilon_p-\frac{\kappa}{2}\alpha_{sj}^2\right)dt, \label{eq:dalpha_pj} \\
d\beta_{pj} 
=&\left(-\gamma_p\beta_{pj}+\epsilon_p-\frac{\kappa}{2}\beta_{sj}^2\right)dt
\end{align}
and for the injection-path modes,
\begin{align}
d\alpha_{cjl} 
=& \left[-\gamma_c\alpha_{cjl} - \zeta_{jl}\left(\alpha_{sj}-\alpha_{sl}e^{-i\theta_{jl}}\right)\right]dt, \\
d\beta_{cjl}
=& \left[-\gamma_c\beta_{cjl} - \zeta_{jl}\left(\beta_{sj}-\beta_{sl}e^{i\theta_{jl}}\right)\right]dt \label{eq:dbeta_cjl}
\end{align}
for $j<l$.
We assume that the pump and injection-path modes decay much faster than the signal modes, i.e., $\gamma_p, \gamma_c \gg \gamma_s$.
The pump and injection-path modes are thus adiabatically eliminated.
Note that we have simplified the model more than the two-DOPO model previously investigated~\cite{K.Takata2015} 
in which the injection-path mode was not eliminated.
Substituting the values for instantaneous steady states of the modes into Eqs.~(\ref{eq:dalphasj}) and (\ref{eq:dbetasj}), 
we obtain 
\begin{align}
\begin{split}
d\mu_j 
= \Bigg[ & -\mu_j + p \nu_j\left(1-\mu_j^2\right) -\xi \left(-\sum_{l(\neq j)}J_{jl}\mu_l - h_j\right) \Bigg]d\tau \\
&+g\sqrt{1-\mu_j^2}dW_{\mu_j}(\tau)
\end{split} \label{eq:dmu_j} \\
\begin{split}
d\nu_j 
= \Bigg[ & -\nu_j + p\mu_j\left(1-\nu_j^2\right) - \xi \left(-\sum_{l(\neq j)}J_{jl}\nu_l - h_j\right) \Bigg]d\tau \\
&+g\sqrt{1-\nu_j^2}dW_{\nu_j}(\tau).
\end{split} \label{eq:dnu_j}
\end{align}
Here we introduced the normalized variables $\mu_j = g\alpha_{sj}/\sqrt{p}$ and $\nu_j=g\beta_{sj}/\sqrt{p}$, 
where $g = \kappa^2/(2\gamma'_s\gamma_p)$ controls the strength of the noise, 
and $p = \kappa\epsilon_p/(\gamma'_s\gamma_p)$ is the pump rate.
We set $p>0$ and do not vary $p$ in time.
The strength of the injection is controlled by $\xi=\xi_0/(\gamma'_s\gamma_c)$.
The parameters $J_{jl} = \zeta_{jl}^2 e^{-i\theta_{jl}}/\xi_0$ and $h_j = g\gamma_c\epsilon_{sj}/(\sqrt{p}\xi_0)$ represent the coupling constant and the longitudinal field, respectively, 
in the Ising Hamiltonian for problems 
which the CIM tries to solve. 
We have set $\zeta_{lj} = \zeta_{jl}$ and $e^{i\theta_{lj}} = e^{-i\theta_{jl}}$. 
To guarantee $J_{jl}$ is real, $e^{-i\theta_{jl}}$ is usually set to $1$ or $-1$.
The parameter $\gamma'_s = \gamma_s + \sum_{l(\neq j)}\zeta_{jl}^2/\gamma_c$ characterizes the effective signal loss and specifies the time scale $\tau = \gamma'_s t$.
To uniquely determine $\gamma'_s$, 
we restrict the setting to satisfy that $\sum_{l(\neq j)}\zeta_{jl}^2$ does not depend on $j$.
By this restriction, 
$J_{jl}$ lies mainly in two classes.
One contains the couplings for the regular graph with uniform magnitude.
This class includes the ferromagnetic Ising model on a lattice.
The system with the coupling $J_{jl} = J$ or $-J$ on a lattice is also included.
The sum $\sum_{j(\neq l)}|J_{jl}|$ in this class is equal to $zJ$, 
where $z$ denotes the coordination number.
The other class is that a site $j$ connects a large number $O(N)$ of sites, 
and $J_{jl}$ is determined by an independent, identical distribution.
An exmple is the fully-connected Ising spin-glass model, the so-called Sherrington-Kirkpatrick model~\cite{D.Sherrington1975}, 
where $J_{jl}$ is extracted from the Gaussian distribution, independently, identically.
The sum $\sum_{j(\neq l)}|J_{jl}|$ in this class in the large-$N$ limit is almost surely equal to some constant that does not depend on $j$.

We obtain the Fokker-Planck equation of the reduced distribution $\tilde{P}(\bm{\mu},\bm{\nu})$ for the signal modes from the stochastic differential equations under the adiabatical elimination of the other modes, 
\begin{equation}
\begin{split}
\frac{d\tilde{P}(\bm{\mu},\bm{\nu})}{dt} 
= \mathcal{L} \tilde{P} & (\bm{\mu},\bm{\nu}) \\
= \sum_{j=1}^N & \bigg\{ \frac{\partial}{\partial \mu_j} \left[ \mu_j - p\nu_j\left( 1-\mu_j^2\right) +\xi V_{\mu,j} \right] \\
&+ \frac{\partial}{\partial \nu_j} \left[ \nu_j - p\mu_j\left( 1-\nu_j^2\right) +\xi V_{\nu,j} \right] \\
& + \frac{1}{2}\frac{\partial^2}{\partial \mu_j^2}g^2\left(1-\mu_j^2\right) \\
& + \frac{1}{2}\frac{\partial^2}{\partial \nu_j^2}g^2\left(1-\nu_j^2\right) \bigg\} \tilde{P}(\bm{\mu},\bm{\nu}),
\end{split} \label{eq:FP}
\end{equation}
where $V_{\mu,j} = -\sum_{l(\neq j)}J_{jl}\mu_l - h_j$ and $V_{\nu,j} = -\sum_{l(\neq j)}J_{jl}\nu_l - h_j$.
This equation is rewritten as
\begin{equation}
\frac{d\tilde{P}(\bm{\mu},\bm{\nu})}{dt} 
= -\sum_{j=1}^N\left( \frac{\partial S_{\mu_j}}{\partial \mu_j} + \frac{\partial S_{\nu_j}}{\partial \nu_j}\right).
\end{equation}
Here $S_{\mu_j}$ and $S_{\nu_j}$ are given as
\begin{align}
\begin{split}
S_{\mu_j} 
= \bigg[ -\mu_j +p\nu_j & \left( 1-\mu_j^2\right) - \xi V_{\mu,j} \\
&- \frac{1}{2}\frac{\partial}{\partial \mu_j}g^2\left(1-\mu_j^2\right)\bigg] \tilde{P}(\bm{\mu},\bm{\nu}), 
\end{split} \\
\begin{split}
S_{\nu_j} 
= \bigg[ -\nu_j +p\mu_j & \left( 1-\nu_j^2\right) - \xi V_{\nu,j} \\
& - \frac{1}{2}\frac{\partial}{\partial \nu_j}g^2\left(1-\nu_j^2\right)\bigg] \tilde{P}(\bm{\mu},\bm{\nu}),
\end{split}
\end{align}
which compose the probability current.

\section{Steady state distribution}

We derive the stationary solution $\tilde{P}_\text{SS}(\bm{\mu},\bm{\nu})$ of Eq.~(\ref{eq:FP}) that satisfies $\mathcal{L}\tilde{P}_\text{SS}(\bm{\mu},\bm{\nu})=0$.
The fact that the distribution for the positive $P$ representation can be real and positive~\cite{P.Drummond1980Jan, H.Carmichael2002, H.Carmichael2008} allows us to introduce a potential function $\Phi(\bm{\mu},\bm{\nu})$ as 
\begin{equation}
\tilde{P}_\text{SS}(\bm{\mu},\bm{\nu}) = Z_N^{-1} e^{-\Phi(\bm{\mu},\bm{\nu})},
\end{equation}
where $Z_N^{-1}$ is a constant for normalization, referred to as the partition function later.
The probability current for $\tilde{P}_\text{SS}(\bm{\mu},\bm{\nu})$ is expressed as
\begin{align}
\begin{split}
S_{\mu_j} 
= \bigg[ &-\left(1-g^2\right)\mu_j +p\nu_j\left( 1-\mu_j^2\right) - \xi V_{\mu,j} \\
&+ \frac{1}{2}g^2\left(1-\mu_j^2\right)\frac{\partial \Phi}{\partial \mu_j} \bigg]\tilde{P}_\text{SS}(\bm{\mu},\bm{\nu}),
\end{split} \\
\begin{split}
S_{\nu_j} 
= \bigg[ &-\left(1-g^2\right)\nu_j +p\mu_j\left( 1-\nu_j^2\right) - \xi V_{\nu,j} \\
& + \frac{1}{2}g^2\left(1-\nu_j^2\right)\frac{\partial\Phi}{\partial \nu_j}\bigg]\tilde{P}_\text{SS}(\bm{\mu},\bm{\nu}).
\end{split}
\end{align}

A simple strategy to find a solution is to assume detailed balance 
that guarantees the existence of the equilibrium distribution as a stationary distribution~\cite{H.Risken1989}.
The detailed balance condition in the Fokker-Planck equation is equivalent to the absence of the probability current~\cite{H.Risken1989}, 
$S_{\mu_j} = S_{\nu_j} = 0 \ \forall j$.
The potential function under the detailed balance condition, say $\Phi_\text{DB}(\bm{\mu},\bm{\nu})$, satisfies
\begin{align}
\frac{\partial \Phi_\text{DB}}{\partial \mu_j}
&= \frac{2}{g^2\left(1-\mu_j^2\right)}\left[\left(1-g^2\right)\mu_j - p\nu_j\left( 1-\mu_j^2\right) + \xi V_{\mu,j}\right], \label{eq:dPhi_DBdmu_j} \\
\frac{\partial \Phi_\text{DB}}{\partial \nu_j}
&= \frac{2}{g^2\left(1-\nu_j^2\right)}\left[\left(1-g^2\right)\nu_j - p\mu_j\left( 1-\nu_j^2\right) + \xi V_{\nu,j}\right]. \label{eq:dPhi_DBdnu_j}
\end{align}
For $\xi \neq 0$, however, no function satisfies the above eqautions, 
since the above equations lead to 
\begin{align}
\frac{\partial}{\partial\mu_l}\frac{\partial\Phi_\text{DB}}{\partial \mu_j} 
&= -\frac{2\xi J_{jl}}{g^2\left(1-\mu_j^2\right)} \label{eq:second-derivative_Phi_DB_jl} \\
\frac{\partial}{\partial\mu_j}\frac{\partial\Phi_\text{DB}}{\partial \mu_l} 
&= -\frac{2\xi J_{jl}}{g^2\left(1-\mu_l^2\right)}, \label{eq:second-derivative_Phi_DB_lj}
\end{align}
where we have used $J_{lj} = J_{jl}$.
The differentiation of the function with respect to $\mu_j$ and $\mu_l$ thus cannot be interchanged for $\xi \neq 0$.
This fact demonstrates that the state distribution governed by Eq.~(\ref{eq:FP}) for $\xi \neq 0$ does not exhibit detailed balance. 
Hence, there exists a finite probability current, 
and the system has no equilibrium state distributions.
When $\xi = 0$, 
where the DOPOs do not interact with each other, 
the detailed balance can be holded.
On the other hand, 
if the noises in the stochastic differential equations [Eqs.~(\ref{eq:dmu_j}) and (\ref{eq:dnu_j})] were governed by the Gaussian distribution, 
i.e., $gdW_{\mu_j}(\tau)$ and $gdW_{\mu_j}(\tau)$, 
the denominators in the right hand side of Eqs.~(\ref{eq:second-derivative_Phi_DB_jl}) and (\ref{eq:second-derivative_Phi_DB_lj}) would turn to $g^2$, 
and the detailed balance could be recovered.
The breakdown of detailed balance thus stems from combination of the interactions and multiplicative noises.

The violation of the detailed balance condition has been discussed in several studies, 
in particular, in the context of efficient sampling methods, 
where it has been confirmed that faster convergence to steady states is realized by violating the detailed balance condition~\cite{H.Suwa2010, K.Turitsyn2011, H.Fernandes2011, Y.Sakai2013}.
In other words, 
the stochastic dynamics without detailed balance can reach the steady states in a shorter time 
than the corresponding dynamics 
that obeys the detailed balance condition~\cite{A.Ichiki2013, M.Ohzeki2015Jul, M.Ohzeki2015Sep}.
The violation of the detailed balance condition of the dynamics in the CIM is hence expected to accelerate the convergence to the steady states.
Note that the origin of the noises is quantum fluctuations.

While the breakdown of the detailed balance condition suggests a nontrivial character of the CIM in the relaxation to steady states, 
it is an obstacle that makes it difficult to derive steady state distributions.
We here focus on a part of the steady state distribution in which 
the difference in the magnitude $\mu_i^2$ ($\nu_j^2$) of different DOPOs is small.
We represent
\begin{equation}
\mu_j^2 = q^\mu + \delta_j^\mu, \ \ \ 
\nu_j^2 = q^\nu + \delta_j^\nu,
\end{equation}
where
\begin{equation}
q^\mu = \frac{1}{N}\sum_{j=1}^N \mu_j^2, \ \ \ 
q^\nu = \frac{1}{N}\sum_{j=1}^N \nu_j^2.
\end{equation}
In a region, 
where $\delta_j^\mu$ and $\delta_j^\nu$ are small, 
there can be a potential function that approximately satisfies Eqs.~(\ref{eq:dPhi_DBdmu_j}) and (\ref{eq:dPhi_DBdnu_j}).
Note again that 
the potential function of the CIM in general never exhibits detailed balance, 
but there could be a part in which Eqs.~(\ref{eq:dPhi_DBdmu_j}) and (\ref{eq:dPhi_DBdnu_j}) hold.
If the other part of the potential function is nearly equal to zero, 
we do not have to consider that part.
The expansion of the potential function in terms of $\delta$, 
obtained from Eqs.~(\ref{eq:dPhi_DBdmu_j}) and (\ref{eq:dPhi_DBdnu_j}), 
is
\begin{equation}
\begin{split}
g^2\Phi 
&=  g^2\Phi_0 -\frac{2\xi}{1-q^\mu}\sum_{j<l}J_{jl}\mu_j \mu_l -\frac{2\xi}{1-q^\nu}\sum_{j<l}J_{jl}\nu_j \nu_l \\
& -\frac{2\xi}{1-q^\mu}\sum_{j=1}^N h_j \mu_j -\frac{2\xi}{1-q^\nu}\sum_{j=1}^N h_j \nu_j + O(\delta^2), \label{eq:g^2Phi}
\end{split}
\end{equation}
where
\begin{equation}
\begin{split}
g^2\Phi_0 
= -N\left(1-g^2\right) [ \ln \left(1-q^\mu \right) + & \ln \left( 1-q^\nu \right) ] \\
& - 2p\sum_{j=1}^N \mu_j \nu_j.
\end{split} \label{eq:Phi_0}
\end{equation}
The function $\Phi_0$ provides the terms for the independent DOPOs, 
reproducing the known result for a single DOPO~\cite{M.Wolinsky1988}, 
which with large $p$ form a double-well in the potential function.
The double-well corresponds to the bifurcation of the in-phase amplitude above the threshold of pump rate 
to make pseudo Ising variables.
The other terms undertake the coupling of DOPOs embedded for a target optimization problem.
When we neglect terms $O(\delta^2)$,
the terms for the coupling compose of the Hamiltonian or the cost function for the continuous relaxation of a target discrete optimization problem.
In the subspace that satisfies $\delta = 0$ the Ising Hamiltonian is effectively reproduced~\cite{T.Leleu2017}.
In the view of steady state distributions, 
the CIM thus finds solutions with the combination of the Ising-like double-wells and the continuous relaxation of the problem.
It is not obvious, however, that the global minimum of the potential function given by Eq.~(\ref{eq:g^2Phi}) agrees with the ground state of the Ising Hamiltonian for the discrete optimization problem.
Note that the presence of the double-well does not directly indicate the superposition of the two coherent states for the wells, namely our up-spin and down-spin.
It rather leads to the classical mixture of them in the single- and two-DOPO cases at least~\cite{M.Reid1992, K.Takata2015}.
In time for transient evolution, a sign of the superposition was found in numerical simulations of a single DOPO case~\cite{L.Krippner1994} and in the two-DOPO model~\cite{K.Takata2015}.
This feature might be a characteristic property of DOPOs even for solving combinatorial optimization problems, 
but the transient time scale is out of our scope in the present study.

\section{Typical solutions with the potential function in the large-size limit}

If the higher order terms $O(\delta^2)$ in Eq.~(\ref{eq:g^2Phi}) are negligible, 
the potential function under the detailed balance condition can be a good approximation of the true one.
It is available to analytically evaluate the property of the approximate potential function without those terms.
We here examine the potential function without the higher order terms, 
applying it to two simple examples.
We consider only the real part of $\mu_j$ and $\nu_j$.
This simplification is based on the standard initial condition of the dynamics in which all signal fields is set to the vacuum state 
and on real $J_{jl}$ with $e^{-i\theta_{jl}} = 1, -1$.
Equations~(\ref{eq:dmu_j}) and (\ref{eq:dnu_j}) show that
if all $\mu_j$ and $\nu_j$ have no imaginary part at an instantaneous time,
they remain real for all time~\cite{M.Wolinsky1988}.

What we like to know is the configuration of Ising spins yielded from the CIM according to the potential function.
The Ising spins are encoded in the sign of the in-phase amplitude of the signal fields~\cite{T.Inagaki2016Oct, P.McMahon2016, Z.Wang2013, A.Marandi2014, K.Takata2015}.
We define an operator $\hat{\sigma}_j$ by $\hat{\sigma}_j \ket{x_j} = \text{sign}(x_j)\ket{x_j}$, 
where $\ket{x_j}$ is the eigenstate of operator $\hat{x}_j = (\hat{a}_{sj} + \hat{a}_{sj}^\dagger)/2$, 
and sign($x_j$) is 1 if $x_j>0$ and $-1$ if $x_j<0$.
Its expectation value is
\begin{equation}
\begin{split}
\text{tr} & \left( \hat{\rho}\hat{\sigma}_j \right) \\
&= \int d^N\bm{x} d^N\bm{\mu}d^N\bm{\nu} \text{sign}\left(x_j\right) \frac{\langle\bm{x}|\bm{\alpha}\rangle\langle\bm{\beta}^*|\bm{x}\rangle}{\langle\bm{\beta}^*|\bm{\alpha}\rangle} \tilde{P}\left(\bm{\mu}, \bm{\nu} \right) \\
&= \int d^N\bm{\mu}d^N\bm{\nu} \left\{ 1 - 2H\left[\frac{\sqrt{p}}{g}\left(\mu_j + \nu_j \right)\right]\right\} \tilde{P}\left(\bm{\mu}, \bm{\nu} \right) \\
&\simeq \int d^N\bm{\mu}d^N\bm{\nu} \text{sign}\left(\mu_j + \nu_j \right) \tilde{P}\left(\bm{\mu}, \bm{\nu} \right), 
\end{split}
\end{equation}
where $H(x) = \int_x^\infty dt e^{-t^2/2}/\sqrt{2\pi}$.
The last line is obtained from the second line by ignoring the fluctuation in the coherent state.
This is a rather good approximation for small $g$.

For estimating the efficiency of the CIM to solve combinatorial optimization problems, 
it is important to invesigate its behavior for large-size problems.
The method of statistical mechanics is suitable for this situation~\cite{T.Aonishi2017}.
We define the partition function and free energy for the potential function by
\begin{align}
Z_N(\eta) 
&= \int d^N\bm{\mu}d^N\bm{\nu} e^{-\Phi(\bm{\mu},\bm{\nu})-g^{-2}\eta M(\bm{\sigma})}, \\
f(\eta) 
&= -\lim_{N\to\infty} \frac{g^2}{N}\ln Z_N(\eta),
\end{align}
repectively.
Here the term $g^2\eta M(\bm{\sigma})$ is introduced to evaluate the expectation value of order parameter $M(\bm{\sigma})$, e.g., $M(\bm{\sigma}) = \sum_{j=1}^N \sigma_j$, where $\sigma_j = \text{sign}(\mu_j + \nu_j)$.
If we had the true solution $\bm{\sigma}^{0}$ of the problem, 
$M(\bm{\sigma})$ could be overlap between the Ising spins in the CIM and the solution, i.e., 
$M(\bm{\sigma}) = \sum_{j=1}^N \sigma_j\sigma_j^0$, 
which estimates how correct the answer of the CIM is.
The free energy chracterizes the macroscopic property of the system, 
giving the expectation values of macroscopic quantities, e.g., $M(\bm{\sigma})$.

\subsection{Fully-connected ferromagnetic coupling}

We first investigate, as the simplest example, an optimization problem
that is mapped onto the fully-connected ferromagnetic Ising model without the Zeeman terms.
All the coupling constants $J_{jl}$ are equal to $J/(2N)$, $J>0$, 
and the longitudinal field $h_{j}$ vanishes.
The correct ground states of the corresponding Ising model are all-up and all-down.
The potential function without the higher order terms $O(\delta^2)$ for this problem is
\begin{equation}
g^2\Phi 
= g^2\Phi_0 -\frac{N\xi J}{2\left(1-q^\mu \right)} \left(m^\mu\right)^2 -\frac{N\xi J}{2\left(1-q^\nu \right)} \left( m^\nu \right)^2, \label{eq:Phi_ferro}
\end{equation}
where $\Phi_0$ is given by Eq.~(\ref{eq:Phi_0}), and
\begin{equation}
m^\mu = \frac{1}{N} \sum_{j=1}^N \mu_j, \ \ \ 
m^\nu = \frac{1}{N} \sum_{j=1}^N \nu_j. \label{eq:m^mu-m^nu}
\end{equation}
The order paramter is $M(\bm{\sigma}) = \sum_{j=1}^N \sigma_j$.

The partition function is written as
\begin{equation}
\begin{split}
Z(\eta)
=& \int d^N\bm{\mu} d^N\bm{\nu} d^2\bm{m}d^2\bm{q} \delta\left(Nm^\mu - \sum_{j=1}^N\mu_j \right) \\
&\times \delta\left(Nm^\nu - \sum_{j=1}^N\nu_j \right) \delta\left(Nq^\mu - \sum_{j=1}^N\mu_j^2 \right) \\
&\times \delta\left(Nq^\nu - \sum_{j=1}^N\nu_j^2 \right) e^{-\Phi-g^{-2}\eta M(\bm{\sigma})} \\
=& \int d^2\bm{m}d^2\bm{q}d^2\bm{\tilde{m}}d^2\bm{\tilde{q}} \exp \bigg( g^{-2}N \\
& \times \bigg\{ \left(1-g^2\right) \left[ \ln \left(1-q^\mu\right) + \ln \left(1-q^\nu\right) \right] \\ 
& \hspace{20pt} + \bm{\tilde{m}}^\text{T} \bm{m} + \bm{\tilde{q}}^\text{T}\bm{q} \\
& \hspace{20pt} + \frac{\xi J}{2\left(1-q^\mu \right)}\left(m^\mu\right)^2 + \frac{\xi J}{2\left(1-q^\nu \right)}\left(m^\nu\right)^2 \\
& \hspace{20pt} + g^2 \ln \int d\mu d\nu e^{-g^{-2}\phi}
\bigg\} \bigg).
\end{split} \label{eq:Z_ferro}
\end{equation}
Here $\bm{m} = (m^\mu,m^\nu)^\text{T}$, $\bm{q} = (q^\mu,q^\nu)^\text{T}$, $\bm{\tilde{m}} = (\tilde{m}^\mu,\tilde{m}^\nu)^\text{T}$, and $\bm{\tilde{q}} = (\tilde{q}^\mu,\tilde{q}^\nu)^\text{T}$, 
where $\bm{x}^\text{T}$ denotes the tranpose of a column vector $\bm{x}$.
The variables $\bm{\tilde{m}}$ and $\bm{\tilde{q}}$ are introduced for the integral expression of the delta function.
In addition, we have
\begin{equation}
\phi = \bm{z}^\text{T}\tilde{Q}\bm{z}+ \bm{\tilde{m}}^\text{T}\bm{z} + \eta \sigma, \label{eq:phi}
\end{equation}
where $\bm{z} = (\mu, \nu)^\text{T}$, 
and elements of matrix $\tilde{Q}$ are $\tilde{Q}_{11} = \tilde{q}^\mu$, $\tilde{Q}_{12} = \tilde{Q}_{21} = -p$, and $\tilde{Q}_{22} = \tilde{q}^\nu$, 
and $\sigma = \text{sign}(\mu + \nu)$.
The integral is calculated as
\begin{equation}
\begin{split}
&\int d\mu d\nu e^{-g^{-2}\phi} \\
&= \frac{\pi g^2}{\sqrt{\text{det}\tilde{Q}}} \exp\left(\frac{g^{-2}}{4}\bm{\tilde{m}}^\text{T}\tilde{Q}^{-1}\bm{\tilde{m}} \right) G\left(\bm{\tilde{m}}, \bm{\tilde{q}}, \eta \right).
\end{split} \label{eq:integral_phi}
\end{equation}
We do not explicitly show $G\left(\bm{\tilde{m}}, \bm{\tilde{q}}, \eta \right)$, 
but the function for the symmetric case discussed later is given in App.~\ref{app:G}.
It should be noted that $G\left(\bm{\tilde{m}}, \bm{\tilde{q}}, 0 \right) = 1$.
Since the exponent of the integrand of $Z(\eta)$ is propotional to $N$, 
in the large $N$ limit the method of steepest descent gives
\begin{equation}
\begin{split}
f(\eta) 
= & \underset{\bm{m},\bm{q},\bm{\tilde{m}},\bm{\tilde{q}}}{\text{extr}} 
\bigg\{ -\left(1-g^2\right)\left[ \ln \left(1-q^\mu \right) + \ln \left(1-q^\nu \right) \right] \\
& - \bm{\tilde{m}}^\text{T}\bm{m} - \bm{\tilde{q}}^\text{T}\bm{q} \\
&- \frac{\xi J}{2\left(1-q^\mu \right)}\left(m^\mu\right)^2 - \frac{\xi J}{2\left(1-q^\nu \right)}\left(m^\nu\right)^2 \\
& - g^2\ln\left(\pi g^2\right) + \frac{g^2}{2}\ln\text{det}\tilde{Q} - \frac{1}{4}\bm{\tilde{m}}^\text{T}\tilde{Q}^{-1}\bm{\tilde{m}} \\
& -g^2\ln G\left(\bm{\tilde{m}}, \bm{\tilde{q}}, \eta \right) \bigg\},
\end{split}
\end{equation}
where $\text{extr}_{\bm{m},\bm{q},\bm{\tilde{m}},\bm{\tilde{q}}}$ represents taking an extremum with respect to $\bm{m},\bm{q},\bm{\tilde{m}},\bm{\tilde{q}}$.
The terms in the forth line contribute to the entropic part of the free energy.
Since the term in the last line is only used to compute the average of $\sigma$,
$\eta$ is set to zero 
when searching for saddle points of the free energy. 
The saddle points are determined as
\begin{align}
\tilde{m}^\mu 
&= -\frac{\xi J}{1-q^\mu}m^\mu, \\
\tilde{q}^\mu 
&= \frac{1-g^2}{1-q^\mu} - \frac{\xi J}{2\left(1-q^\mu \right)^2}\left(m^\mu \right)^2, \\
m^\mu
&= -\frac{\tilde{m}^\mu \tilde{q}^\nu + \tilde{m}^\nu p}{2\left( \tilde{q}^\mu \tilde{q}^\nu - p^2 \right)}, \\
q^\mu
&= \frac{g^2\tilde{q}^\nu}{2\left( \tilde{q}^\mu \tilde{q}^\nu - p^2 \right)} + \left[ \frac{\tilde{m}^\mu \tilde{q}^\nu + \tilde{m}^\nu p}{2\left( \tilde{q}^\mu \tilde{q}^\nu - p^2 \right)} \right]^2.
\end{align}
We also have the equations obtained by interchanging $\mu$ and $\nu$ in superscripts in the above equations.
In particular, $\bm{m}$ and $\bm{q}$ for the saddle points are the expectation values of them under the distribution governed by the potential function in Eq.~(\ref{eq:Phi_ferro}).

We have not found any solutions that satisfy $m^\mu \neq m^\nu$ by numerical calculations.
We thus restrict ourselves to consider symmetric solutions for which the parameters do not depend on $\mu$ and $\nu$, i.e., 
$m^\mu = m^\nu = m$, $q^\mu = q^\nu = q$, $\tilde{m}^\mu = \tilde{m}^\nu = \tilde{m}$, $\tilde{q}^\mu = \tilde{q}^\nu = \tilde{q}$.
Accordingly, the saddle point equations reduce to
\begin{align}
\tilde{m}
&= -\frac{\xi J}{1-q} m, \label{eq:tildem^sp} \\
\tilde{q} 
&= \frac{1-g^2}{1-q} - \frac{\xi J}{2\left(1-q \right)^2} m^2, \\
m
&= -\frac{\tilde{m}}{2\left( \tilde{q} - p \right)}, \\
q
&= \frac{g^2\tilde{q}}{2\left( \tilde{q}^2 - p^2 \right)} + \left[ \frac{\tilde{m}}{2\left( \tilde{q} - p \right)} \right]^2. \label{eq:q^sp}
\end{align}
Intuitively, 
the term $-2g^{-2}p\sum_{j}\mu_j \nu_j$ in the potential function $\Phi_0$ [Eq.~\ref{eq:Phi_0}] enhances the overlap between $\mu$ and $\nu$, 
and then the symmetric solution is realized.
It should be noted that the restriction that 
the macroscopic parameters above do not depend on $\mu$ and $\nu$ 
does not mean that 
we assume $\mu_j = \nu_j$.
The condition $\mu_j = \nu_j$ would restrict our analysis into a smaller subspace, 
where the density operator is represented as a classical mixture of the coherent states.
The condition, consequently, leads to a different entropic part from the above one we actually obtained.
This difference indicates that our analysis including approximations still reflects some quantum effects.

What we like to calculate is the expectation value of the Ising spins $m_\sigma = \langle N^{-1} \sum_{j=1}^N \sigma_j \rangle$, 
where $\langle X \rangle$ denotes the average of $X$ over the distribution $e^{-\Phi}Z(\eta=0)^{-1}$.
Using the function $G$ in the symmetric case shown in App.~\ref{app:G}, 
we obtain 
\begin{equation}
m_\sigma 
= \frac{df(\eta)}{d\eta}\bigg|_{\eta=0} 
= -1 + 2H\left( \frac{\tilde{m}}{g\sqrt{\tilde{q}-p}} \right), \label{eq:m_sigma}
\end{equation}
where the values of parameters $\tilde{m}$ and $\tilde{q}$ are for the saddle point.

We first examine the solution for $\xi = 0$, 
where the system has no interactions between different DOPOs.
In this case, 
the saddle point equations give $m= m_\sigma = 0$ 
that is consistent with the fact that the system has no bias.
The solution $q$ has a positive value for $g>0$.
In the limit $g \to 0$, 
$q$ shows not the fluctuation but just the square of the amplitude of $\mu$ and $\nu$ frozen at a basin of the potential function.
Thus we can find the character of the shape of the potential function in the behavior of $q$.
There is a threshold $p = 1$.
Below the threshold, $p < 1$, 
the solution of $q$ is equal to zero, 
which means that 
the potential function has the unique minimum at $\mu = \nu= 0$.
Above the threshold, $p > 1$, 
another solution appears with a finite value, $q = 1 - 1/p$, in addition to $q=0$.
The potential function then have the minima at $\mu = \nu = \pm \sqrt{1-1/p}$ and the unstable extremum at $\mu = \nu = 0$.
This behavior agrees with the known bifurcation for a single DOPO~\cite{M.Wolinsky1988}.

We move to investigation of the system for finite $\xi$, 
where the DOPOs interact with each other.
To gain insight into this case, 
we consider the limit $g \to 0$, 
where Eq.~(\ref{eq:q^sp}) reduces to $q = m^2$.
Accordingly, 
the saddle-point equation for $m$ results in 
\begin{equation}
\begin{split}
m =& \frac{1}{2}\left[\frac{1}{1-m^2} - \frac{\xi J}{2\left(1-m^2\right)^2}m^2 - p\right]^{-1} \\
& \times \frac{\xi J}{1-m^2}m 
\ \ \ (g=0).
\end{split}
\end{equation}
This equation has three (five) possible solutions of $m^2$ ($m$);
$m_0 = 0$ and $m_\pm^2 = 1-(1\pm\sqrt{1-2p\xi J})/2p$.
To choose physically reasonable solutions, 
we examine the stability of the possible ones.
The first candidate $m_0$ is stable only when $p+\xi J/2<1$, 
and the others are unstable or complex in this condition.
The second one $m_+^2$ is stable only when $p > 1/2$ and $p+\xi J/2>1$, 
but the region $2p\xi J>1$ is excluded, 
where $m_+$ becomes complex.
The third one $m_-^2$ is always unstable or complex.
Summarizing, 
we have a finite real solution, $m_+$, only for $p > 1/2$ and $p+\xi J/2>1$ except for $2p\xi J>1$.
When $p+\xi J/2<1$, the solution is $m_0 (=0)$.
In the other region our approach does not yield any real solution.
These solutions determine $m_\sigma$ via Eq.~(\ref{eq:m_sigma}); 
negative (positive) $\tilde{m}$, i.e., positive (negative) $m$, leads to $m_\sigma = 1$ ($m_\sigma = -1$), 
which is the correct ground state of the corresponding Ising model.
The boundary, therefore, is $p + \xi J/2 = 1$, 
and larger $p$ and $\xi$ under $p >1/2$ give $m_\sigma = 1$ or $-1$~\cite{T.Aonishi2017}.
When $g=0$, $p + \xi J_{ij}N$ is an effective pump rate, 
if all the DOPOs display the same $\mu$ and $\nu$.
In this example, where $J_{ij} = J/2N$, 
$p + \xi J/2$ is the effective pump rate.
The boundary obtained here is given by the effective pump rate equal to unity. 

The emergence of the finite solution is identified with a phase transtion in the $p$--$\xi J$ phase space.
If we carry out the annealing approach by controling $p$ or $\xi$ with keeping the above instantaneous steady states, 
the system undergoes the phase transition.
It is interesting that 
$m_\sigma$ exhibits the discontinuous change at the boundary, 
namely, the first-order phase transition, 
while $m$ continuously changes as the second-order one.
This definite difference is only for the case $g \to 0$, 
but this finding suggests a feature of the scheme, 
in which the problem is solved by the continous variables, i.e., $\mu$ and $\nu$, 
encoding the discrete ones.

\begin{figure}
	\begin{center}
	\includegraphics[width=\columnwidth]{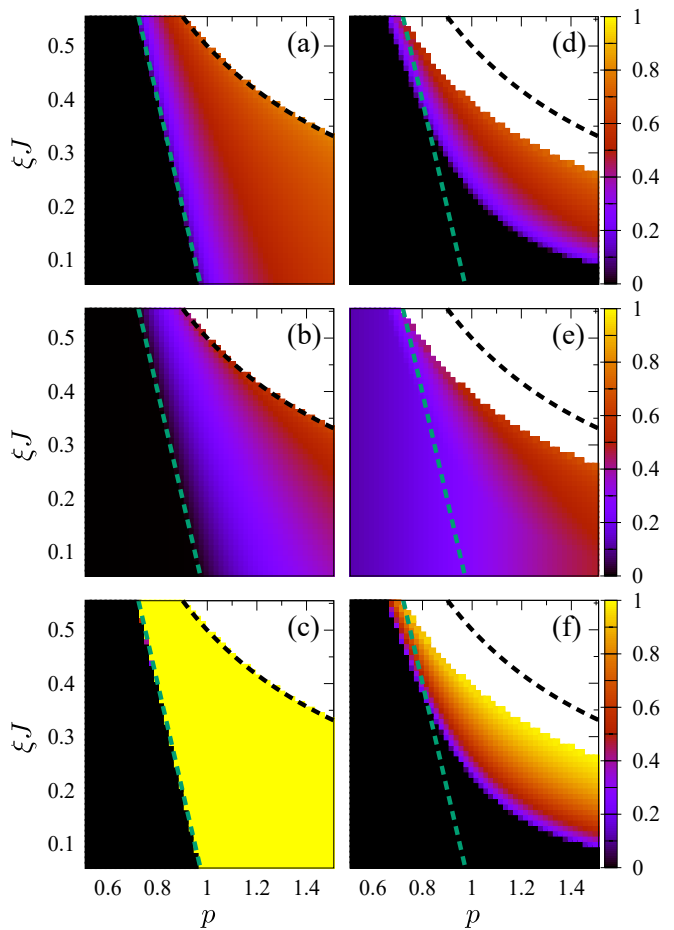}
	\end{center}
	\caption{Heatmap for the solutions of the saddle-point equations [Eqs.~(\ref{eq:tildem^sp})--(\ref{eq:q^sp})] for different $g$ in the $p$--$\xi J$ plane.
		(a) $m$, (b) $q$, and (c) $m_\sigma$ for $g=0.01$, 
		and (d) $m$, (e) $q$, and (f) $m_\sigma$ for $g=0.4$.
		The dashed straight line is the boundary above which $m$ for $g \to 0$ has a finite real value, 
		while the dashed curve is the boundary above which there is no real, stable solution of $m$ for $g \to 0$.
		Initial conditions in solving the equations are set for simplicity so that $m$ tends not to be negative.
		The white region is where the real, stable solutions are not found.
	}
	\label{fig:solutions^sp_ferro}
\end{figure}

We numerically find the stable solutions of the saddle-point equations for $g>0$.
For small finite $g$ ($=0.01$) [Fig.~\ref{fig:solutions^sp_ferro}~(a)--(c)], 
the solution is similar to that for $g=0$.
Note that $g$ for the actual CIM is smaller than 0.01.
We find a rather sharp transition of $m_\sigma$ from 0 to a finite value very close to unity, 
at the almost same location as that for $g = 0$.
The solution $m_\sigma = -1$ is also obtained, 
but we set an initial condition in solving the equations so that $m$ tends not to be negative for simplicity.
The result for a rather large $g$ ($=0.4$) [Fig.~\ref{fig:solutions^sp_ferro}~(d)--(f)] displays different behavior due to noises.
The boundary for $m$ to take a finite value becomes a curve instead of the straight line as for $g=0$.
A stronger pump is thus needed to have finite $m$.
This is because $g$ partially plays a role of temperature in the dynamics of our system.
Hence, the system with large $g$ has large fluctuations.
The value of $m$ thus tends to vanish, 
whereas $q$ takes a finite value.
However, 
it is only when $p$ and $\xi$ is small that 
the system is completely disturbed by the fluctuations.
Stronger pump and interactions make basins in the potential function deeper.
The state of the system is captured by the deep basin.
A finite $m$ is observed, consequently.
Whereas $m_\sigma$ for $g=0.01$ suddenly changes at the boundary, 
its change for $g=0.4$ is smoother due to the noises.
We nevertheless are able to obtain almost all-up (or all-down) state with larger $p$ and $\xi$.

\subsection{Fully-connected ferromagnetic coupling with random fields}

We next examine the fully-connected ferromagnetic coupling with random fields. 
The couplings are the same as those in the previous example, 
$J_{ij} = J/2N$, $J>0$, 
while a longitudinal field for each site is randomly taken from some distribution identically, independently.
The potential function without the higher order terms $O(\delta^2)$ for this problem is
\begin{equation}
\begin{split}
g^2\Phi 
=& g^2\Phi_0 -\frac{N\xi J}{2\left(1-q^\mu \right)} \left(m^\mu\right)^2 -\frac{N\xi J}{2\left(1-q^\nu \right)} \left( m^\nu \right)^2 \\
& -\frac{2\xi}{1-q^\mu}\sum_{j=1}^N h_j \mu_j -\frac{2\xi}{1-q^\nu}\sum_{j=1}^N h_j \nu_j,
\end{split}
\end{equation}
where $\Phi_0$ and both $m^\mu$ and $m^\nu$ are given in Eqs.~(\ref{eq:Phi_0}) and (\ref{eq:m^mu-m^nu}), respectively.
We also set $M(\bm{\sigma}) = \sum_{j=1}^N \sigma_j$ in this problem.
We examine the case in which 
the fields take binary values $h_0$ or $-h_0$ at random.
The corresponding Ising model governed by the Gibbs distribution has been investigated well~\cite{H.Nishimori2011}.
It is known that the system at zero temperature undergoes a first-order phase transition 
with increasing the amplitude $h_0$ of the fields.
The field for the transition is $h_0/J = 1/2$, 
below which the spins are all-up or all-down, 
whereas each spin is parallel to the field above the critical point.
Our aim here is to clarify whether our approach captures this transition.

It is not difficult, 
as shown in App.~\ref{app:random-field}, 
to extend the partition function and free energy for the no-field model to the case with fields, 
but the randomness in the fields has to be carefully treated.
We here exploit the self-averaging property, 
where the free energy for an instance of random fields is almost surely equal to the averaged one in the large $N$ limit~\cite{H.Nishimori2011}.
We can derive the corresponding saddle-point equations for the symmetric solutions from the averaged free energy over the configuration of random fields,
\begin{align}
\tilde{m}
&= -\frac{\xi J}{1-q} m, \label{eq:tildem^sp_random-field} \\
\tilde{q} 
&= \frac{1-g^2}{1-q} - \frac{\xi J}{2\left(1-q \right)^2} m^2 - \frac{2\left(\xi h_0 \right)^2}{\left(\tilde{q} - p\right)\left(1-q\right)^3}, \label{eq:tildeq^sp_random-field} \\
m
&= -\frac{\tilde{m}}{2\left( \tilde{q} - p \right)}, \label{eq:m^sp_random-field} \\
q
&= \frac{g^2\tilde{q}}{2\left( \tilde{q}^2 - p^2 \right)} + \frac{\tilde{m}^2 + \left[2\xi h_0/\left(1-q\right)\right]^2}{4\left( \tilde{q} - p \right)^2}. \label{eq:q^sp_random-field}
\end{align}
The expectation value of Ising spins is
\begin{equation}
\begin{split}
m_\sigma =& -1 + H\left[\frac{\tilde{m}-2\xi h_0/(1-q)}{g\sqrt{\tilde{q}-p}}\right] \\
& + H\left[\frac{\tilde{m}+2\xi h_0/(1-q)}{g\sqrt{\tilde{q}-p}}\right]. 
\end{split} \label{eq:m_sigma_random-field}
\end{equation}

In the limit $g\to 0$, 
the saddle-point equations for finite $m$ lead to $q = m^2 + q_h$, 
where $q_h = (2h_0/J)^2$ indicates the variance of $\mu$ and $\nu$ purely driven by the random fields.
Accordingly, Eq.~(\ref{eq:m^sp_random-field}) with Eq.~(\ref{eq:tildeq^sp_random-field}) turns to 
\begin{equation}
\begin{split}
m 
= \frac{1}{2} & \Bigg[\frac{1}{1-m^2-q_h} \\
& - \frac{\xi J}{2\left(1-m^2-q_h\right)^2}\left(m^2+2q_h\right) -p \Bigg]^{-1} \\
& \times \frac{\xi J}{1-m^2-q_h} m \ \ \ (g=0).
\end{split}
\end{equation}
This equation has three (five) possible solutions of $m^2$ ($m$);
$m_0=0$ and $m_\pm^2 = (1-q_h)[1-(1\pm \sqrt{1-2p'\xi'J})/(2p')]$, 
where $p' = (1-q_h)p$ and $\xi' = (1+q_h)\xi/(1-q_h)$.
We can find the physical solutions, 
which are real and stable ones, 
through the same argument as in the no-field case, 
but $p$ and $\xi$ in that case are replaced with $p'$ and $\xi'$ here.
We hence have a finite solution ($m_+$) 
only when $p' > 1/2 $ and $ p' + \xi' J/2 > 1$, 
but the region $2p'\xi'J>1$, 
where $m_+$ becomes complex, 
is excluded.
For $p' + \xi'J/2 <1$, 
the solution is $m_0(=0)$.
In the other region,
$p'<1/2 $ and $ p' + \xi'J/2 >1$, 
there is no real solution.
These solutions determine $m_\sigma$ via Eq.~(\ref{eq:m_sigma_random-field}) with Eq.~(\ref{eq:tildem^sp_random-field});
if $m^2 > q_h$, $|m_\sigma| = 1$, 
otherwise $m_\sigma = 0$.
The discontinuous change of $|m_\sigma|$ from 0 to $1$ thus occurs at $h_0/(m_+ J) = 1/2$ in the region for $p' > 1/2 $ and $ p' + \xi' J/2 > 1$.
The quantity $h_0/m_+$ represents the magnitude of the effective field in the Ising model 
which the CIM actually solves, 
since being divided by $m_+$ relaxes the discrepancy, 
in the balance between the two-body and one-body interactions, 
of the model represented with $\mu$ and $\nu$ from the Ising model.
This finding demonstrates that
the CIM detects the first-order phase transition at which 
the ratio of the effective field to the coupling constant is equal to 1/2.
This boundary agrees with that for the corresponding Ising system at zero temperature governed by the conventional Gibbs distribution~\cite{H.Nishimori2011}.

When we compare the condition $h_0/(m_+J) = 1/2$ with the transition point in the corresponding Ising model, 
it is considered as the boundary for $h_0$ with fixed $p$ and $\xi$ in the region for finite $m$.
The condition is also interpreted as the boundary for $p$ or $\xi$ with fixed $h_0$.
For the latter we discuss the transition which the system undergoes in the annealing approach with controling $p$ or $\xi$.
As in the no-field case, 
$m$ continuously changes with increasing $p$ or $\xi$, 
while $m_\sigma$ jumps at the boundary.
There is a difference, however, in the mechanism for the jump.
The discontinous change in the random-field case is caused by the first-order phase transition of the genuine Ising model, 
whereas that observed in the no-field case is just due to the bifurcation of the DOPOs.
The first-order phase transition is in general owed to the presence of an energy barrier between multiple minima~\cite{H.Nishimori2011}, 
which makes it difficult to search for the ground states in the energy landscape.
Hence, that transition should be avoided in the scheme.
For instance, 
such a transition in quantum annealing often concerns an exponentially small energy gap~\cite{T.Jorg2008, T.Jorg2010Feb, T.Jorg2010May, C.Laumann2012} 
and thus inefficienty of the method.
The continuous change of $m$ in our approach demonstrates the absence of local minima, 
despite the target Ising model has the first-order phase transition.

\begin{figure}
	\begin{center}
	\includegraphics[width=\columnwidth]{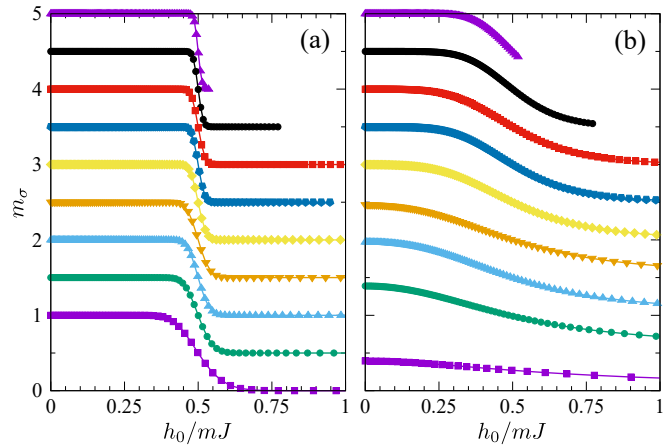}
	\end{center}
	\caption{Obtained $m_\sigma$ from Eq.~(\ref{eq:m_sigma_random-field}) 
		with the solution of the saddle-point equations [Eqs.~(\ref{eq:tildem^sp_random-field})--(\ref{eq:q^sp_random-field})] 
		as a function of $h_0/(mJ)$ 
		for (a) $g= 0.01$ and (b) $g=0.1$ 
		at $(p, \xi) = (1.0,0.1)$, $(1.1, 0.1)$, $(1.2, 0.1)$, $(1.0, 0.2)$, $(1.1, 0.2)$, $(1.2, 0.2)$, $(1.0, 0.3)$, $(1.1, 0.3)$, and $(1.2, 0.3)$ 
		from bottom to top.
		The plots are shifted vertically by 0.5 for clarity.
		Real, stable solutions at large $h_0/mJ$ are not obtained for some sets of $(p,\xi)$.
		Initial conditions in solving the saddle-point equations are set for simplicity so that $m$ tends not to be negative.
	}
	\label{fig:m_sigma_random-field}
\end{figure}

We find the solutions of the saddle-point equations for $g>0$ by numerically solving them.
For $g=0.01$ the resulting $m_\sigma$ steeply decreases from 1 to 0 with increasing $h_0/(mJ)$ with fixed $p$ and $\xi$ as shown in Fig.~\ref{fig:m_sigma_random-field}~(a).
The sudden change takes place around $h_0/(mJ) = 0.5$, 
which agrees with the transition point of the corresponding Ising model at zero temperature.
The transition yielded by our model, however, is not the first-order one, 
while the corresponding Ising model exhibits the first-order one even for finite low temperatures~\cite{H.Nishimori2011}.
This difference probably originates from the fact that 
our model is governed by the continuous degrees of freedom, $\mu$ and $\nu$, 
rather than the discrete ones in the genuine Ising model.
Larger $p$ and $\xi$ enhance Ising-like behavior.
As a result, steeper change at the transition point is found.
Except around the transition point, 
we obtain the correct ground state of the target Ising model.
For larger $g$ [Fig.~\ref{fig:m_sigma_random-field}~(b)], 
we do not find any sudden change of $m_\sigma$, 
but it monotonically decreases with increasing the effective field.

\section{Summary and discussion}

We investigated network of degenerate optical parametric oscillators (DOPOs) 
as a model of the coherent Ising machine (CIM)~\cite{Z.Wang2013, A.Marandi2014, T.Inagaki2016Oct, P.McMahon2016}, 
an architecture for solving problems expressed with the Ising models.
The network is composed of the optical coupling of DOPOs, 
representing the Ising model with the parameters $J_{jl}$ and $h_j$.
Motivated by the annealing approach~\cite{S.Geman1984, H.Nishimori1998, T.Albash2018, L.Venuti2016}, 
we tried to derive the steady state distributioins for the dynamics described with the positive $P$ representation~\cite{P.Drummond1980Jan}.
The distribution is expected to yield answers of the CIM on the problems in the long time limit.
We obtained approximate steady state distributions for arbitrary Ising problems under the ansatz that 
the difference in $\mu_i^2$ and $\nu_i^2$ from other DOPOs is small.
Using the method of statistical mechanics in the large problem-size limit, 
we showed a prescription to obtain the most probable states in the distributions 
in which higher order terms for the inhomogeneity of $\mu_i^2$ and $\nu_i^2$ are neglected.
For two rather simple problems, 
i.e., fully-connected ferromagnetic coupling without/with binary random fields, 
we derived phase diagrams in the $p$--$\xi J$ plane.
The phase diagrams demonstrate that 
the most probable states in a particular range of the parameters correspond to the true optimal states.
In particular, in the random-field problem, 
the distribution correctly detects the phase transition 
that occurs in the genuine Ising model with varying the magnitude of the fields.
We found through this analysis an interesting feature of our system that
despite the nature of the first-order phase transition in terms of the Ising spins is effectively detected, 
the order parameter $m$ for the quadrature amplitude does not show discontinuous change.

Our analysis is based on the approximate steady state distributions, 
but it is probably valid for the no-field problem, 
since the problem has no effect 
that disturbs the uniformity in the magnitudes.
On the other hand, 
the random fields contribute to the growth of the difference in the magnitude.
Hence, 
our result of the random-field problem, in particular, for the fields with large amplitude might be modified 
when including the higher order terms for the inhomogeneity of the magnitude.
Numerical simulations for this problem should be done to examine this issue.
In addition, it is interesting to estimate effects of the higher order terms to the most probable states.

The obstacle that makes it difficult to derive the exact steady state distributions is the violation of the detailed-balance condition in the dynamics.
This is not just an obstacle but a property 
that has us expect faster convegence to steady states than 
the conventional dynamics fluctuated by the simple Gaussian noise~\cite{H.Suwa2010, K.Turitsyn2011, H.Fernandes2011, Y.Sakai2013, A.Ichiki2013, M.Ohzeki2015Jul, M.Ohzeki2015Sep}.
The violation stems from the multiple facts;
the multiplicative noise and the coupling of the DOPOs.
The former, generated by the coupling of the signal and pump fields, 
also appears in a single DOPO, 
but it is not enough for the breakdown of the detailed balance.
The coupling of the DOPOs is thus essential to the peculiar dynamics. 
The dynamics of the DOPOs with this property might provide an advantage of the CIM in solving problems.
Investigation on this subject is left as a future work.

\appendix

\section{Function $G\left(\tilde{m}, \tilde{q}, \eta \right)$}
\label{app:G}

We restrict ourselves to consider the symmetric case, 
where $\tilde{m}^\mu = \tilde{m}^\nu = \tilde{m}$ and $\tilde{q}^\mu = \tilde{q}^\nu = \tilde{q}$.
The function $\phi$ in Eq.~(\ref{eq:phi}), after diagonalizing $\tilde{Q}$, is rewritten as
\begin{equation}
\phi = \tilde{q}^+ w_1^2 + \tilde{q}^- w_2^2 - \frac{1}{4}\bm{\tilde{m}}^\text{T}\tilde{Q}^{-1}\bm{\tilde{m}} + \eta \text{sign}\left(\sqrt{2}w_2 - \frac{\tilde{m}}{\tilde{q}^-}\right).
\end{equation}
Here we have used eigenvalues $\tilde{q}^\pm$ of $\tilde{Q}$, 
which for the symmetric case are $\tilde{q}^\pm = \tilde{q} \pm p$, and
\begin{equation}
\begin{pmatrix}
w_1 \\
w_2 
\end{pmatrix}
= \bm{w} = V\left(\bm{z} + \frac{1}{2}\tilde{Q}^{-1}\bm{\tilde{m}} \right), 
\ \ \ 
V = \frac{1}{\sqrt{2}}
\begin{pmatrix}
1 & -1 \\
1 & 1 
\end{pmatrix}
.
\end{equation}
The function $\text{sign}(x)$ gives $1$ if $x>0$ and $-1$ if $x<0$.
We then calculate the integral in Eq.~(\ref{eq:integral_phi}) for the symmetric case as
\begin{equation}
\begin{split}
\int d\mu d\nu & e^{-g^{-2}\phi} \\
=& e^{ \frac{g^{-2}}{4}\bm{\tilde{m}}^\text{T}\tilde{Q}^{-1}\bm{\tilde{m}} } 
\int_{-\infty}^{\infty} dw_1 e^{-g^{-2}\tilde{q}^+w_1^2} \\
& \times \Bigg[ \int_{-\infty}^{\tilde{m}/\left(\sqrt{2}\tilde{q}^-\right)} dw_2 e^{-g^{-2}\left( \tilde{q}^-w_2^2 -\eta \right)} \\ 
&+ \int_{\tilde{m}/\left(\sqrt{2}\tilde{q}^-\right)}^\infty dw_2 e^{-g^{-2}\left( \tilde{q}^-w_2^2 + \eta \right)} \Bigg] \\
=& \frac{\pi g^2}{\sqrt{\tilde{q}^+ \tilde{q}^-}} \exp\left(\frac{g^{-2}}{4}\bm{\tilde{m}}^\text{T}\tilde{Q}^{-1}\bm{\tilde{m}} \right) \\
& \times \left[ e^{g^{-2}\eta} - 2 \sinh \left( g^{-2}\eta \right) H\left( \frac{\tilde{m}}{g\sqrt{\tilde{q}^-}} \right) \right], 
\end{split}
\end{equation}
where $H(x) = \int_x^\infty dt e^{-t^2/2}/\sqrt{2\pi}$.
Comparing this with Eq.~(\ref{eq:integral_phi}), 
we obtain
\begin{equation}
G\left( \tilde{m}, \tilde{q}, \eta \right) 
= e^{g^{-2}\eta} - 2 \sinh \left( g^{-2}\eta \right) H\left( \frac{\tilde{m}}{g\sqrt{\tilde{q}^-}} \right). \label{eq:G}
\end{equation}
The function $G(\bm{\tilde{m}}, \bm{\tilde{q}}, \eta)$ without the symmetry is also obtained through similar calculations.


\section{Free energy for the random-field case}
\label{app:random-field}

The partition function for the random-field case is same as that for the no-field case [Eq.~(\ref{eq:Z_ferro})], 
except that $\ln \int d\mu d\nu e^{-g^{-2}\phi}$ in the latter is replaced with $N^{-1}\sum_{j=1}^N \ln \int d\mu d\nu e^{-g^{-2}\phi^\text{RF}_j}$ in the former.
Here we have
\begin{equation}
\phi^\text{RF}_j 
= \bm{z}^\text{T}\tilde{Q}\bm{z}+ \bm{\tilde{m}}_{h_j}^\text{T}\bm{z} + \eta \sigma,
\end{equation}
and
\begin{equation}
\tilde{m}^\mu_{h_j} = \tilde{m}^\mu - \frac{2\xi h_j}{1-q^\mu},
\ \ \ 
\tilde{m}^\nu_{h_j} = \tilde{m}^\nu - \frac{2\xi h_j}{1-q^\nu}, 
\end{equation}
for $\bm{\tilde{m}}^\text{T}_j = (\tilde{m}^\mu_{h_j}, \tilde{m}^\nu_{h_j})$.
The function $\phi^\text{RF}_j$ has the similar form to $\phi$ [Eq.~(\ref{eq:phi})], 
and the integral $\int d\mu d\nu e^{-g^{-2}\phi^\text{RF}_j}$ is calculated as in the no-field case.
The sum of logarithm of the obtained function concerns only fields.
The sum in the large $N$ limit hence corresponds to the average with respect to the random field, 
showing the self-averaging property~\cite{H.Nishimori2011}.
We thus have
\begin{equation}
\begin{split}
&\frac{1}{N}\sum_{j=1}^N \ln \int d\mu d\nu e^{-g^{-2}\phi^\text{RF}_j} \\
&= \ln \frac{\pi g^2}{\sqrt{\text{det}\tilde{Q}}} + \frac{g^{-2}}{4} \left\langle \bm{\tilde{m}}_h^\text{T}\tilde{Q}^{-1}\bm{\tilde{m}}_h \right\rangle_h + \left\langle \ln G\left(\bm{\tilde{m}}_h, \bm{\tilde{q}}, \eta \right) \right\rangle_h, 
\end{split}
\end{equation}
where $\langle X \rangle_h$ denotes the average of $X$ over the random field $h$.
We then obtain, 
by the method of steepest descent, 
the free energy for the random-field case, 
\begin{equation}
\begin{split}
f^\text{RF}(\eta)
= & \underset{\bm{m},\bm{q},\bm{\tilde{m}},\bm{\tilde{q}}}{\text{extr}} 
\bigg\{ -\left(1-g^2\right)\left[ \ln \left(1-q^\mu \right) + \ln \left(1-q^\nu \right) \right] \\
& - \bm{\tilde{m}}^\text{T}\bm{m} - \bm{\tilde{q}}^\text{T}\bm{q} \\
& - \frac{\xi J}{2\left(1-q^\mu \right)}\left(m^\mu\right)^2 - \frac{\xi J}{2\left(1-q^\nu \right)}\left(m^\nu\right)^2 \\
& - g^2\ln\left(\pi g^2\right) + \frac{g^2}{2}\ln\text{det}\tilde{Q} - \frac{1}{4} \left\langle \bm{\tilde{m}}_h^\text{T}\tilde{Q}^{-1}\bm{\tilde{m}}_h \right\rangle_h \\
& -g^2 \left\langle \ln G\left(\bm{\tilde{m}}_h, \bm{\tilde{q}}, \eta \right) \right\rangle_h \bigg\}.
\end{split}
\end{equation}
Let us consider random fields each of which takes either $h_0$ or $-h_0$ at random.
For the symmetric case, 
where $\tilde{m}^\mu = \tilde{m}^\nu = \tilde{m}$ and $\tilde{q}^\mu = \tilde{q}^\nu = \tilde{q}$, 
the average in the free energy results in
\begin{align}
\left\langle \bm{\tilde{m}}_h^\text{T}\tilde{Q}^{-1}\bm{\tilde{m}}_h \right\rangle_h 
=& \frac{2}{\tilde{q} - p} \left[ \tilde{m}^2 + \left( \frac{2\xi h_0}{1-q} \right)^2 \right], \\
\begin{split}
\left\langle \ln G\left(\tilde{m}_h, \tilde{q}, \eta \right) \right\rangle_h
= & \frac{1}{2} \ln G\left(\tilde{m}-\frac{2\xi h_0}{1-q}, \tilde{q}, \eta \right) \\
&+ \frac{1}{2} \ln G\left(\tilde{m}+\frac{2\xi h_0}{1-q}, \tilde{q}, \eta \right), 
\end{split}
\end{align}
where $G\left(\tilde{m}, \tilde{q}, \eta \right)$ is given in Eq.~(\ref{eq:G}).

\begin{acknowledgments}
The authors thank A. Ichiki and K. Ohki for helpful discussions and comments on the manuscript.
R.~M. thanks T. Leleu and Y. Yamamoto for valuable discussions.
This research is supported by the Impulsing Paradigm Change Through Disruptive Technologies (ImPACT) Program of the Council of Science, 
Technology and Innovation (Cabinet Office, Government of Japan).
M.~O. acknowledges JSPS KAKENHI No.~15H03699, No.~16H04382, and No.~16K13849.
\end{acknowledgments}

\bibliography{ref_cim_steady-states}

\end{document}